# Direct Imaging, Three-dimensional Interaction Spectroscopy, and Friction Anisotropy of Atomic-scale Ripples on MoS$_2$

Omur E. Dagdeviren[1,#], Ogulcan Acikgoz[2,#], Peter Grütter[1] and Mehmet Z. Baykara[2,*]

[1]Department of Physics, McGill University, Montréal, Québec H3A 2T8, Canada

[2]Department of Mechanical Engineering, University of California Merced, Merced, CA 95343, USA

# These authors contributed equally to this work.

* e-mail: mehmet.baykara@ucmerced.edu

## Abstract

Theory predicts that two-dimensional (2D) materials may only exist in the presence of out-of-plane deformations on atomic length scales, frequently referred to as *ripples*. While such ripples can be detected via electron microscopy, their direct observation via surface-based techniques and characterization in terms of interaction forces and energies remain limited, preventing an unambiguous study of their effect on mechanical characteristics, including but not limited to friction anisotropy. Here, we employ high-resolution atomic force microscopy to demonstrate the presence of atomic-scale ripples on supported samples of few-layer molybdenum disulfide (MoS$_2$). Three-dimensional force / energy spectroscopy is utilized to study the effect of ripples on the interaction landscape. Friction force microscopy reveals multiple symmetries for friction anisotropy, explained by studying rippled sample areas as a function of scan size. Our experiments contribute to the continuing development of a rigorous understanding of the nanoscale mechanics of 2D materials.

**Introduction**

The discovery that atomically thin sheets can be isolated from bulk crystals and the exciting physical properties exhibited by them, initiated the thriving field of two-dimensional (2D) materials[1]. Over a span of more than 15 years, the electrical, mechanical, and chemical properties of 2D materials – including graphene, molybdenum disulfide ($MoS_2$) and others – were studied in great detail, revealing extraordinary characteristics that could eventually allow revolutionary applications in diverse areas of science and technology[2-5].

Despite the prevalent scientific interest in 2D materials today, their very existence initially puzzled scientists, based on the idea that a perfectly two-dimensional crystalline sheet of material would be thermodynamically unstable[6]. Subsequent work revealed that such 2D materials may in fact exist in the presence of out-of-plane deformations with atomic-scale (< 1 nm) corrugations, also termed *ripples*[7,8]. The presence of ripples was confirmed by transmission electron microscopy (TEM) imaging performed on suspended single-layer graphene[7], as well as single-layer $MoS_2$[9], although in an indirect fashion, i.e. by studying the broadening of diffraction spots in reciprocal space. Eventually, TEM experiments also allowed the direct visualization of atomic-scale ripples in suspended few-layer (up to ~10 layers) graphene samples in real space, with out-of-plane corrugations and lateral spacings on the order of 5 Å and 500 Å, respectively, largely in accordance with theoretical expectations[10]. Perhaps more importantly from an application point of view, scanning tunneling microscopy (STM) experiments demonstrated that the rippling of graphene is preserved even for samples that are supported on substrates such as silicon dioxide ($SiO_2$)[11], in an *intrinsic* fashion that is not related to the topographical features of the substrate itself.

Most prior work related to ripples focused on their effect on the electrical properties of 2D materials. For instance, the presence of ripples was found to suppress weak localization in graphene[12] and attempts were made to control the structure and distribution of ripples in order to tune electrical properties[13,14]. On the other hand, the number of studies focusing on the effect of ripples on the mechanical characteristics of 2D materials is much lower. A particular mechanical phenomenon that was studied in detail on 2D materials is friction, based on the discovery that single- or few-layers of graphene, $MoS_2$ and other 2D materials function as effective solid lubricants on the nanoscale, where lubrication with fluids is impractical due to size effects[15]. The pioneering work on 2D material friction conducted via atomic force microscopy (AFM)[15] was soon followed by other AFM experiments on graphene that demonstrated a strong dependency of friction forces on the scanning direction, i.e. *friction anisotropy*[16]. A peculiar aspect of these milestone experiments was that the friction forces exhibited a 2-fold symmetry, in obvious contradiction to the 6-fold symmetry of the atomic structure of graphene. This observation, also made by other groups on other 2D materials including $MoS_2$[17], was explained by the presumed existence of linearly aligned structures (i.e. ripples) on the material surface, leading to high (low) friction forces when the AFM tip is scanning across (along) the ripples on a particular region of the sample. Despite this seemingly widespread idea – together with a competing theory based on the presence of linearly aligned stripes formed by environmental adsorbates[18] –, the connection to friction anisotropy remains controversial, as the ripples are not directly observed during the experiments. Moreover, the literature also includes friction anisotropy studies that deviate from 2-fold symmetry[19,20].

Here, we employ AFM-based high-resolution imaging, three-dimensional force / energy spectroscopy, and nanoscale friction measurements for a detailed physical characterization of atomic-scale ripples on few-layer $MoS_2$ samples supported on $SiO_2$. Our results directly demonstrate the presence of linearly aligned ripples with heights limited to only a few Å on the samples. Three-dimensional force / energy spectroscopy[21,22] is used to go beyond topographical imaging, and quantify the influence of ripples on the interaction energies experienced by the probe tip in close proximity to the sample surface, revealing modulations of the potential energy landscape down to a few meV. Friction force microscopy (FFM) measurements performed at different scan sizes (ranging from a few tens of nm to a few µm laterally) lead to 2-fold and 4-fold friction anisotropies as well as the occasional observation of non-periodic friction anisotropy data. A Fourier-transform-based analysis of lateral force data derived from three-dimensional potential energy maps as a function of scan size provides clues regarding the observed variety in friction anisotropy.

## Results

**Imaging of Atomic-scale Ripples on $MoS_2$**

Motivated by the absence of AFM data in the literature demonstrating the presence of atomic-scale ripples on 2D materials, we performed experiments to answer the question of whether the imaging of atomic-scale ripples on $MoS_2$ can be accomplished with an AFM-based approach. Based on the defining role that ripples are thought to play in the friction anisotropy of such materials[16,17], we initially conducted FFM experiments on few-layer $MoS_2$ samples exfoliated onto $SiO_2$, whereby topographical and friction force maps are recorded simultaneously as the AFM tip slides on the sample surface in contact mode[23]. While the friction force map of Fig. 1a, recorded on a multi-layer $MoS_2$ flake, clearly demonstrates the layer-dependence of friction that is a ubiquitous characteristic of 2D materials[15], no linearly aligned structures, i.e. ripples, are observed on the flake surface, along the lines of previous FFM work performed on this material[15,17,20].

Considering that FFM necessitates continuous contact between the AFM tip and the sample surface, which invariably results in the averaging of tip-sample interactions over a finite contact area and thus leads to a loss of spatial resolution[24], we directed our attention to alternative modes of AFM imaging. In particular, imaging via conventional *tapping-mode* AFM (performed by way of amplitude modulation[25], with oscillation amplitudes on the order of 10 nm) did not result in the imaging of ripples either (Fig 1b). This result is perhaps not surprising considering that tapping-mode AFM, despite the absence of a continuous contact between the tip and the sample, still involves intermittent contact (manifesting in the form of repulsive tip-sample interactions), which also results in a loss of spatial resolution.

In order to overcome the limitations of FFM and tapping-mode AFM in terms of spatial resolution, we imaged the topography of exfoliated $MoS_2$ flakes via frequency-modulation atomic force microscopy (FM-AFM)[26] performed in the attractive tip-sample interaction regime (an approach which is frequently referred to as noncontact AFM, i.e. NC-AFM[27]). The utilization of ultra-sharp probes (see Methods), combined with the fact that tip-sample contact is avoided during the measurements, finally resulted in the direct imaging of ripples on the $MoS_2$ surface (on a flake of ~65 Å height, corresponding to ~10 layers) in the form of linearly aligned, minute undulations in the surface topography, with out-of-plane corrugations of 1 – 5 Å and lateral spacings on the order of 300 – 400 Å (Fig. 1c). These results, which constitute the first direct imaging of atomic-scale ripples on a 2D material such as $MoS_2$, at the same time open up the way for their detailed characterization in the form of interaction forces and energies.

**Three-dimensional Interaction Spectroscopy of Atomic-scale Ripples on $MoS_2$**

While the presence of atomic-scale ripples on the $MoS_2$ samples (Fig. 1c) may initially appear as a purely structural feature, it is important to probe their effect on the interactions that the 2D material exhibits with other bodies in its vicinity, in particular due to the potentially defining role they play in intriguing nanoscale mechanical characteristics such as friction anisotropy uncovered by AFM[16,17].

Motivated by this line of argument, we performed three-dimensional force / energy spectroscopy[21,22] on the $MoS_2$ flake of Fig. 1c to extract the tip-sample interaction landscape in the form of three-dimensional, volumetric maps of interaction energies and forces, with meV- and pN-level resolution, respectively. The data, collected in the form of 106 constant-frequency-shift topography maps at different tip-sample distances via FM-AFM (Fig. 2a), are combined to reconstruct the three-dimensional interaction force / energy volume (see Methods). Subsequently, two-dimensional maps of tip-sample interaction energy at fixed tip-sample distances are extracted from the three-dimensional data (Fig. 2b-f), which allows a high-resolution study of how ripples modulate the tip-sample interaction landscape both spatially and energetically. The analysis of the data reveals that the mean energy corrugation associated with the ripples increases with decreasing tip-sample distance, from 5 meV to 30 meV over nearly 6 nm. This trend, which also points to an increasing magnitude of lateral forces experienced by the AFM tip near the ripples at decreasing tip-sample distances (which are proportional to the lateral gradient of the interaction energy in the scanning direction), demonstrates the non-negligible effect of atomic-scale ripples on mechanical characteristics of $MoS_2$ probed by recording tip-sample interactions in AFM experiments.

**Friction Anisotropy on $MoS_2$**

Considering that the direction dependence, i.e. anisotropy, of friction can be potentially an important design parameter for 2D-material-based solid lubrication in small-scale mechanical systems, we performed FFM measurements to probe friction anisotropy on $MoS_2$ flakes exfoliated onto $SiO_2$. Our work was additionally motivated by previous reports of friction anisotropy on 2D materials including graphene and $MoS_2$, where 2- and 6-fold symmetries have been reported[16-19], as well as irregular anisotropic behavior[20]. While 2-fold anisotropic behavior was tentatively explained by the presence of ripples[16,17] or stripes formed by molecular adsorbates[18], 6-fold anisotropy was ascribed to the hexagonal symmetry of the atomic structure of the involved materials[19].

Analysis of multiple FFM experiments performed in our laboratory, aimed at studying anisotropic friction on $MoS_2$ (for technical details, see Methods), revealed that the results fall into three main categories (Fig. 3): (i) anisotropic behavior with nearly 2-fold symmetry, as demonstrated by data acquired on a large (8 μm × 8 μm) scan area (Fig. 3a), (ii) anisotropic behavior with nearly 4-fold symmetry, as demonstrated by data acquired on a smaller (50 nm × 50 nm) scan area (Fig. 3b), and finally, (iii) non-periodic friction, as demonstrated by data acquired on a scan area of 1.5 μm × 1.5 μm in size (Fig. 3c). No experiments performed on scan areas smaller than 1.5 μm × 1.5 μm featured 2-fold anisotropy, while the largest scan area on which higher-symmetry anisotropy was recorded was 50 nm × 50 nm. Anisotropy ratios (the ratio between the highest and lowest friction values recorded in a ~360° cycle) for the experiments were 2.0 ± 0.4, in the range of previously reported values for graphene and $MoS_2$[16,20].

## Discussion

Despite the fact that the imaging of linearly aligned ripples on $MoS_2$ via our high-resolution AFM experiments can be utilized to explain the widely-reported observation of 2-fold friction anisotropy, the presence of higher-symmetry anisotropies in smaller scan sizes, as well as the observation of non-periodic friction, highlight the need for a more thorough evaluation of the effect of ripples on friction anisotropy.

Motivated as above, we performed a Fourier transform analysis on the two-dimensional lateral force maps derived from the volumetric interaction energy data. In particular, we calculated the relative probability of encountering a type of spatial symmetry (2-, 3-, 4-, 6-fold as well as no symmetry) on areas ranging in size from 16 nm × 16 nm to 250 nm × 250 nm that are scanned over whole lateral force maps (Fig. 4). The results demonstrate that the effect of ripples on lateral force anisotropy (in the form of a 2-fold symmetry) is most dominant at larger scan sizes, while the chances of encountering 2-fold anisotropic behavior rapidly decrease at smaller scan sizes, where the "no symmetry" state dominates.

These results can be understood when one takes the finite lateral size and spacing of the ripples into account, the latter of which is in the range of 100s of Å. As such, in order for the ripples to have a noticeable influence on friction anisotropy, scan sizes need to be relatively larger, a conclusion that is supported by the friction anisotropy experiments reported here, where no experiments performed on scan areas smaller than 1.5 μm × 1.5 μm featured 2-fold anisotropy. This conclusion also shows that the observation of 2-fold friction anisotropy in certain prior reports[16,17] is definitely expected, based on the fact that the reported measurements were performed on areas of multiple μm in lateral size.

On the other hand, it needs to be understood that the predication of "no symmetry" for small scans, delivered by the present Fourier-transform-based analysis, is limited by the fact that the lateral force map analyzed here was acquired over an area of 250 nm × 250 nm. As such, the map lacks atomic-scale spatial resolution and thus cannot capture the effect of the hexagonal symmetry of atomic-scale structure on friction anisotropy. In fact, for FFM data acquired over small areas, it is natural to expect that the hexagonal symmetry of the atomic structure will result in an anisotropic behavior closer to 6-fold symmetry. Consequently, the convolution of this effect with the still non-negligible influence of ripples on scan areas of a few tens of nm in lateral size, results in anisotropic behavior with an intermediate level of symmetry (such as the measurement reported in Fig. 1b that features nearly 4-fold anisotropy). At the other end of the spectrum, once scans are limited only to a few nm in lateral size, the influence of ripples completely disappears, and the emergence of 6-fold anisotropic behavior is expected, as clearly demonstrated by experiments performed on graphene[19].

Despite the fact that the discussion above sheds light on the observation of 2-fold and higher order anisotropic behavior of friction exhibited by 2D materials, the frequent observation of non-periodic friction data (Fig. 1c) needs to be explained, too. While tip apex changes during experiments (that are known to directly affect the magnitude of friction forces during FFM measurements[28]) can be held responsible for such results, we alternatively ascribe the occasional inability to record clearly anisotropic friction data to the fact that the presence and distribution of ripples on the $MoS_2$ flakes appear to be non-uniform, with significant areas on the flakes that are devoid of linearly aligned ripples (Supplementary Fig. 1). As the atomic-scale ripples cannot be imaged during the FFM measurements, it is conceivable that some measurements are ultimately performed on areas with no linearly aligned ripples, resulting in non-periodic results in terms of friction anisotropy.

Finally, it needs to be mentioned that our experiments did not yield any evidence for linearly aligned structures formed by adsorbed molecules as proposed by Gallagher *et al.*[18], even when the $MoS_2$ surfaces were imaged by the high-resolution NC-AFM approach. Along a similar line of thought, in order to rule out that the linearly aligned structures we observe are indeed intrinsic ripples of $MoS_2$ and not clusters of adsorbates on the $MoS_2$ flakes, we studied energy dissipation maps acquired simultaneously with the other data channels during NC-AFM imaging.

The absence of a discernible contrast in such maps (Supplementary Fig. 2) supports the conclusions reached about the nature of the atomic-scale structures we observe as intrinsic ripples.

In summary, we presented high-resolution AFM experiments that led to the direct imaging of atomic-scale ripples on few-layer flakes of $MoS_2$. Three-dimensional force / energy spectroscopy showed the extent to which the presence of the ripples influences the interactions of $MoS_2$ with the probing tip. Our experiments directly revealed the presence of linearly aligned ripples as the fundamental physical mechanism responsible for the direction-dependence of friction on 2D materials, and also allowed the explanation of the wide variety of anisotropic behavior observed on such materials as a function of scan size. Further experiments, potentially performed with probes that are themselves single- or few-layers of 2D materials[29], need to be performed to more accurately ascertain the impact of ripples on 2D-material-based solid lubrication in micro- and nano-scale mechanical systems.

**Methods**

**Friction Force Microscopy.** FFM experiments reported here were performed with a commercial AFM instrument (Asylum Research, Cypher VRS) under ambient conditions. $MoS_2$ flakes were deposited onto $SiO_2$ substrates by standard mechanical exfoliation from commercially available bulk crystals via adhesive tape. The measurements were conducted using diamond-like-carbon-coated and diamond-coated cantilevers (Budget Sensors ContDLC and Nanosensors CDT-CONTR, respectively), with normal spring constant values (0.90 N/m and 0.28 N/m, respectively) as determined by the Sader method in a quite room[30,31]. AFM measurements were performed in contact-mode, whereby the lateral force signal was collected together with topography maps. During the measurements, the effective normal load was purely due to adhesion and the scanning direction was perpendicular to the cantilever main axis. Topography and lateral force maps were acquired at scanning rates of 1 to 2 Hz. In order to investigate friction anisotropy, i.e. record the dependence of friction forces on scanning direction, the sample was manually rotated around its surface normal by ~30° between each measurement shown in Fig. 1, for a full cycle corresponding to ~360°. The measurements focused on areas of a few micrometer-square to a few nanometer-square on few-layer regions of $MoS_2$ flakes (corresponding to less than 10 but more than 4 layers). Each anisotropy experiment (corresponding to a nearly full cycle of friction force measurements reported in each panel of Fig. 3) was completed on the same day, in a continuous experimental run, to minimize variations in tip and sample conditions. Friction force maps were constructed from forward and backward lateral force maps[32], whereby each map consisted of 256 scan lines. Friction force values (Fig. 3b,c) and ratios of friction forces recorded on $SiO_2$ and $MoS_2$ (Fig. 3a), reported for each rotation angle, are extracted from these maps. Specifically, multiple (in particular, four) regions in the corresponding friction maps are considered for each rotation angle; the mean friction force (and friction force ratio) values as well as corresponding standard deviations are derived from these. In order to minimize the potential effect of tip changes during larger scans, the friction force recorded on the $SiO_2$ substrate is used as a reference value and divided by the friction force recorded on $MoS_2$, resulting in friction force ratios (as reported in Fig. 3a).

**Three-dimensional Force / Energy Spectroscopy.** The experiments were conducted using a customized JSPM-5100 microscope. The microscope was equipped with a custom-made sample stage operating in high vacuum (~ $10^{-7}$ mbar). Ultra-sharp gold-coated tips (Adama Innovations AD-2.8-SS, tip radius, $r$ < 5 nm, stiffness, $k$ = 2.0 N/m) were employed and the microscope was controlled with the GXSM control module[33], with the implementation of active drift control. Nanosurf® EasyPLL Plus was used for frequency shift detection. The microscope was operated via the standard frequency modulation atomic force microscopy (FM-AFM) technique with self-excitation[26]. The cantilever was oscillated at its first resonance frequency, $f_0$ = 61,786.3 kHz, with an oscillation amplitude of 10 nm for all experiments. The sensitivity and the noise floor of the cantilever were calibrated with the thermal excitation technique in a quiet room[31]. The measurements were performed on as-exfoliated $MoS_2$ flakes, with no additional preparation under vacuum.

Three-dimensional atomic force microscopy (3D-AFM) is a well-established technique, the details of which can be found elsewhere[21,22]. We implemented constant frequency shift experiments to image the surface topography with different frequency set points. Imaging the same area of the sample at different heights leads to an "imaging volume". We established our imaging volume with 106 layers. Using established methods, we merged all topography and frequency shift data to reconstruct the three-dimensional potential energy landscape of the sample[34], with sub-nm lateral resolution. The lateral force acting on the tip was calculated via the negative gradient of the potential energy along the lateral direction[21]. Similarly, the vertical tip-sample interaction force was recovered with the negative gradient of the potential energy along the vertical direction. We rigorously checked the reconstructed tip-sample interaction data to make sure that it is well-posed[35,36].

**Data Availability.** The data that support the findings of this study are available from the corresponding author upon request.

**Acknowledgements**

This work was supported by the Merced Nanomaterials Center for Energy and Sensing (MACES) via the National Aeronautics and Space Administration (NASA) Grant No. NNX15AQ01. O.E.D. and P.G. acknowledge financial support from the Natural Sciences and Engineering Research Council of Canada and Le Fonds de Recherche du Québec - Nature et Technologies. Banting Fellowship is gratefully acknowledged by O.E.D.

**Author contributions**

M.Z.B conceived the experiments. M.Z.B. and O.E.D. wrote the manuscript. O.E.D performed the three-dimensional interaction spectroscopy experiments. O.A. performed the friction anisotropy experiments. All authors participated in the analysis and interpretation of the data.

**Competing financial interests**

The authors declare that there are no competing interests.

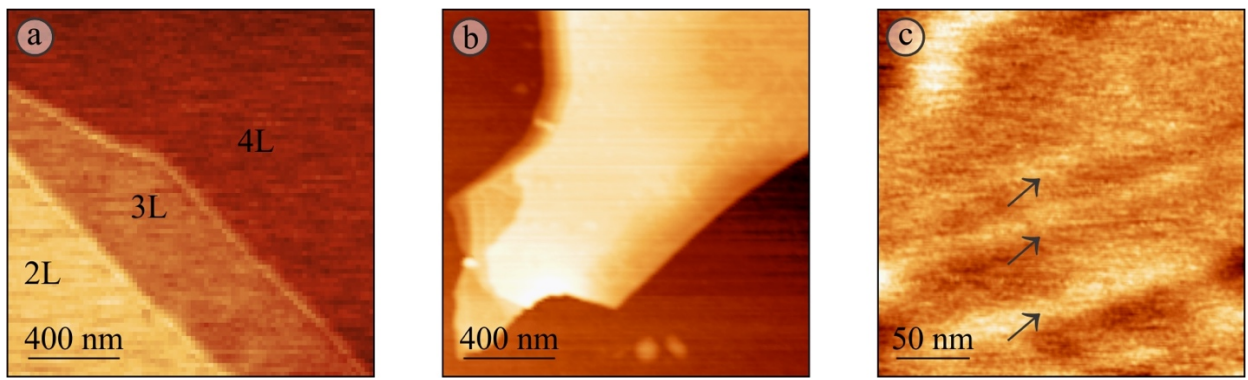

**Figure 1 | Searching for atomic-scale ripples on MoS$_2$ with different AFM techniques. (a)** Friction map acquired on a multi-layer (2L: two layers, 3L: three layers, 4L: four layers) MoS$_2$ flake, where brighter colors correspond to higher friction. While decreasing friction with increasing number of layers is observed (whereby, on average, 3L and 4L exhibit %71 and 50% of the friction recorded on 2L, respectively), no evidence of ripples can be detected. **(b)** Amplitude-modulation (i.e. tapping-mode) atomic force microscopy image of the topography associated with a multi-layer MoS$_2$ flake on a SiO$_2$ substrate (color scale range: 10 nm). No trace of linearly aligned ripples is found on the MoS$_2$ flake. **(c)** Topography image recorded via frequency-modulation atomic force microscopy on a few-layer MoS$_2$ flake, revealing the presence of linearly aligned, atomic-scale ripples on MoS$_2$, highlighted by black arrows (color scale range: 5 Å; also see Supplementary Fig. 1). The height of the highlighted ripples ranges from 1 Å to 3 Å.

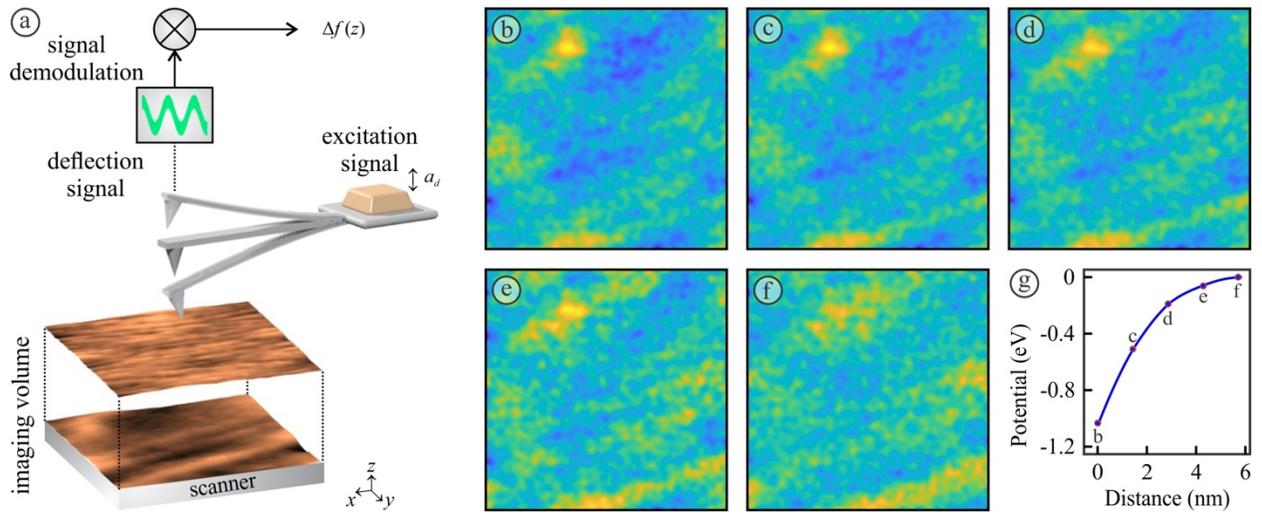

**Figure 2 | Three-dimensional tip-sample interaction spectroscopy of atomic-scale ripples on MoS₂. (a)** The surface topography is imaged at varying tip-sample distances by changing the frequency shift ($\Delta f$) of the cantilever. The oscillation amplitude of the cantilever is kept constant at 10 nm by employing an active feedback of excitation signal, $a_d$. Three-dimensional topography and resonance frequency shift data are utilized to reconstruct tip-sample interaction potential[21,22]. **(b-f)** Maps of tip-sample interaction energy at different tip-sample distances recorded over the same location as in Fig 1c. The average tip-sample interaction energy at each tip-sample distance was subtracted from the data to highlight corrugations. The color scale range decreases from 87 meV in (b) to 7 meV in (f). **(g)** Average tip-sample interaction energy as a function of tip-sample distance. The tip-sample distances and the average tip-sample interaction energies corresponding to the data presented in (b) to (f) are highlighted on the plot.

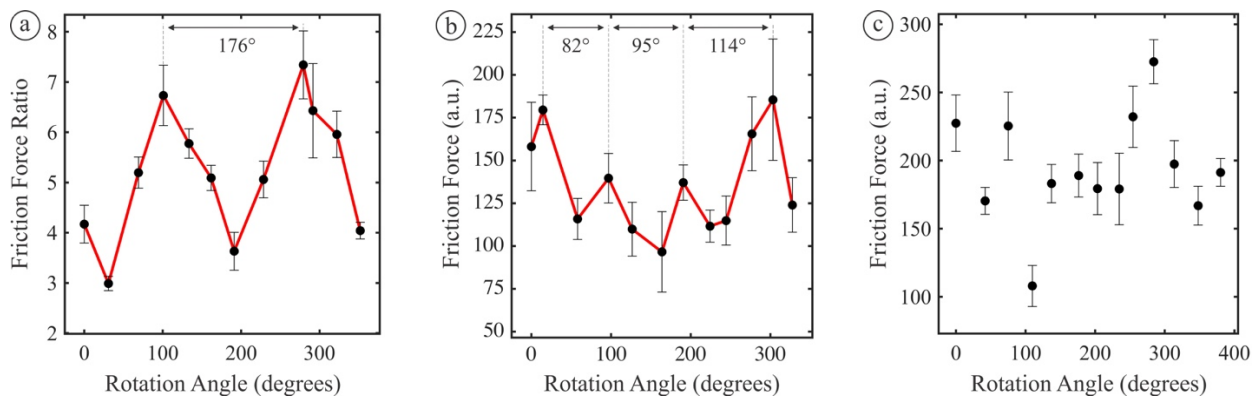

**Figure 3 | Friction anisotropy on MoS$_2$. (a)** Ratios of friction forces (see Methods) recorded on SiO$_2$ and MoS$_2$, as a function of rotation angle, extracted from a scan of 8 μm × 8 μm in size. Periodic behavior with nearly 2-fold symmetry is observed. **(b)** Friction force recorded on MoS$_2$ as a function of rotation angle, extracted from a scan of 50 nm × 50 nm in size. Periodic behavior with nearly 4-fold symmetry is observed. **(c)** Friction force recorded on MoS$_2$ as a function of rotation angle, extracted from a scan of 1.5 μm × 1.5 μm in size, showing non-periodic character.

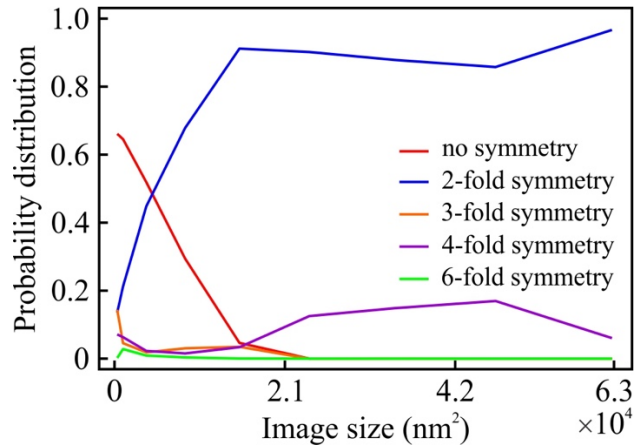

**Figure 4 | Dependence of friction anisotropy on scan size.** The relative probabilities of encountering a particular type of spatial symmetry (2-, 3-, 4-, 6-fold as well as no symmetry at all) on lateral force maps derived from the three-dimensional potential energy data presented in Fig. 2, as a function of scan size (ranging from 256 nm$^2$ to 62,500 nm$^2$). While 2-fold symmetry, due to the presence of ripples, dominates the probability distribution at large scan sizes, the "no symmetry" state is dominant for smaller scans. A non-negligible observation of 4-fold symmetry for large scan sizes can be potentially attributed to the presence of surface structures other than linearly aligned ripples.

# Supplementary Information for

# **Direct Imaging, Three-dimensional Interaction Spectroscopy, and Friction Anisotropy of Atomic-scale Ripples on MoS$_2$**

Omur E. Dagdeviren[1,#], Ogulcan Acikgoz[2,#], Peter Grütter[1] and Mehmet Z. Baykara[2,*]

[1]Department of Physics, McGill University, Montréal, Québec H3A 2T8, Canada

[2]Department of Mechanical Engineering, University of California Merced, Merced, CA 95343, USA

# These authors contributed equally to this work.

* e-mail: <mehmet.baykara@ucmerced.edu>

**Supplementary Figure 1.** Distribution of Ripples

**Supplementary Figure 2.** Energy Dissipation

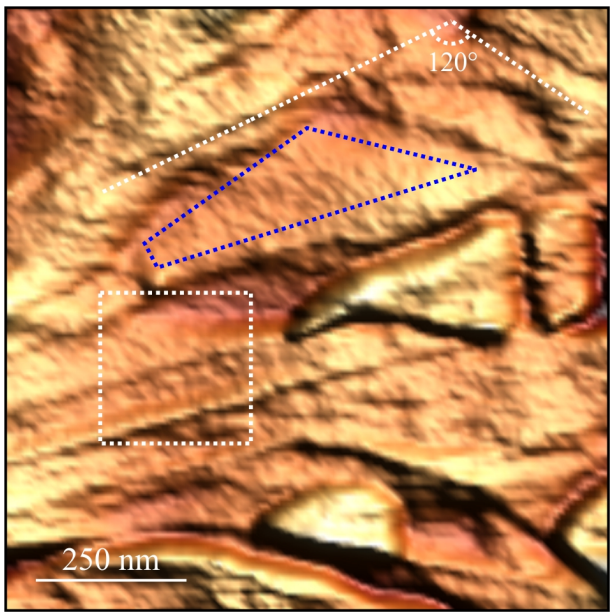

**Supplementary Figure 1 | Distribution of Ripples.** A large-scale (1,000 nm × 1,000 nm) topographical image (in isometric, pseudo-3D form; color scale range: 4.2 nm) of the $MoS_2$ flake acquired via high-resolution FM-AFM, including the region with linearly aligned ripples investigated in Figs. 1 and 2 (dashed white rectangle), a representative region where there are no discernible ripples (area highlighted with the blue dashed lines) and a ripple that changes direction by ~120° (dashed white lines), in accordance with the atomic symmetry of $MoS_2$.

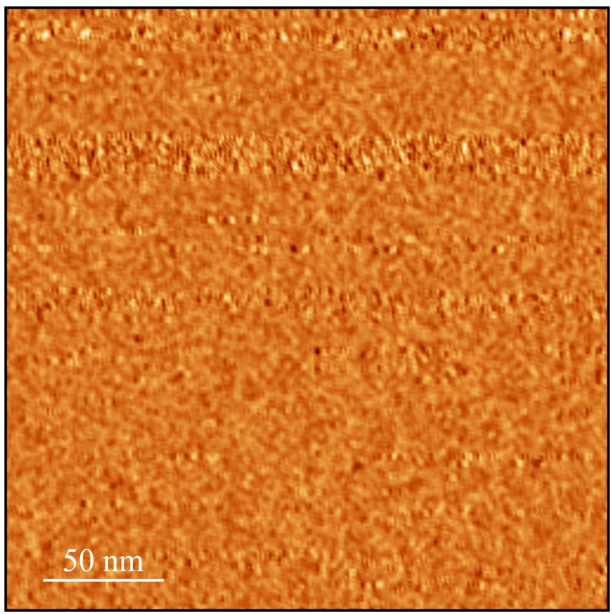

**Supplementary Figure 2 | Energy Dissipation.** A map (250 nm × 250 nm) of energy dissipation recorded simultaneously with the high-resolution topographical image of Fig. 1c (color scale range: 1.8 meV). The lack of contrast demonstrates that the ripples observed in the topography map are indeed inherent structural features and not clusters formed by adsorbates from the environment.